\documentclass{PoS}
\usepackage{amsfonts,amssymb,amsmath,bm}
\usepackage{graphicx}

\newcommand{\be}{\begin{equation}}
\newcommand{\bea}{\begin{eqnarray}}
\newcommand{\ee}{\end{equation}}
\newcommand{\eea}{\end{eqnarray}}
\newcommand{\bpi}{\begin{picture}}
\newcommand{\bce}{\begin{center}}
\newcommand{\epi}{\end{picture}}
\newcommand{\ece}{\end{center}}

\def\chic#1{{\scriptscriptstyle #1}}
\def\diff{{\rm d}}
\def\NV{\bqq'}

\def\NP{ V}

\def\bqq{\mathrm{I}\!\Gamma}

\def\bqq{{\mathrm{I}\!\Gamma}}
\def\ff{B}

\def\bcj{J}
\title{Infrared properties of the gluon mass equation}

\ShortTitle{Infrared properties of the gluon mass equation}

\author{\speaker{Daniele Binosi}\\  
         European Centre for Theoretical Studies in Nuclear Physics
        and Related Areas (ECT*), \\and Fondazione Bruno Kessler,\\
        Villa Tambosi, Strada delle Tabarelle 286, I-38123 Villazzano (TN)  Italy.\\
        E-mail: \email{binosi@ectstar.eu}}

\author{Joannis Papavassiliou\\
        Department of Theoretical Physics and IFIC, University of Valencia-CSIC,\\ 
E-46100, Valencia, Spain.\\}        
\abstract{The gauge-invariant generation of a dynamical, momentum-dependent gluon mass is intimately connected with the 
presence of non-perturbative massless poles in the vertices of the theory,
which trigger the well-known Schwinger mechanism.
In the deep infrared the integral equation that governs 
this effective gluon mass assumes a particularly simple form,
which may be derived following two seemingly different, but ultimately equivalent procedures.
In particular, it may be obtained either as a deviation from 
a special identity that enforces the masslessness of the gluon 
in the absence of massless poles, or as a direct consequence of the 
appearance of a non-vanishing bound-state wave function, associated 
with the details of the actual formation of these massless poles. In this presentation we demonstrate that, 
due to profound relations between the various ingredients,  
the two versions of the gluon mass equation are in fact absolutely identical.}

\FullConference{International Workshop on QCD Green's Functions, Confinement and Phenomenology,\\
		September 05-09, 2011\\
		Trento Italy}

\begin{document}

\section{Introduction}

\noindent      
It is well-established by now that 
the      dynamical  generation of an effective    gluon mass~\cite{Cornwall:1981zr}    
explains in a natural and self-consistent way  
the infrared
finiteness of  the (Landau gauge) gluon propagator  and ghost dressing
function, observed  in large volume lattice  simulations for
both                $SU(2)$~\cite{Cucchieri:2007md}                and
$SU(3)$~\cite{Bogolubsky:2007ud,Bowman:2007du,Oliveira:2009eh}    gauge
groups (for an alternative approach see~\cite{Dudal:2008sp}). Given the non-perturbative nature of the mass generation mechanism, 
the usual starting point in the continuum is the 
Schwinger-Dyson equations (SDEs) governing the 
Green's functions under scrutiny. 
In the framework 
provided      by       the      synthesis      of       the      pinch
technique~(PT)~\cite{Cornwall:1981zr,    Cornwall:1989gv,Binosi:2002ft}
with  the   background  field  method~(BFM)~\cite{Abbott:1980hw}, 
these complicated integral equations are endowed with a variety of 
important properties, which allow a much tighter control 
on the truncations adopted and the approximation schemes employed.

Probably the most crucial theoretical ingredient in this context 
is a special type of vertex, denoted by $V$, 
which contains massless, longitudinally coupled poles.
This vertex complements the all-order three-gluon vertex 
entering into the SDE governing the gluon self-energy,
and captures the underlying mass generating mechanism, which is 
none other than the Schwinger mechanism.
Specifically, the QCD dynamics is assumed to generate  
massless bound-state excitations, which, in turn, give rise to  
the aforementioned poles that appear inside the vertex $V$.

At the level of the SDE for the gluon propagator, 
the analysis finally boils down to the derivation of 
an integral equation that governs the evolution of the 
dynamical gluon mass, $m^2(q^2)$,  
as a function of the momentum $q^2$, in a way similar to 
the more familiar case of 
the dynamical generation of a constituent quark mass. However, 
unlike what happens with the quark gap equation,
which, due to its Dirac structure, is unambiguously separated  
into two equations, governing 
the quark mass and wave-function, the derivation of the corresponding 
equations for the gluon mass and wave-function requires  
the introduction of some 
additional kinematic criteria~\cite{Aguilar:2011ux}.
For the purposes of this presentation, we will focus 
on the integral equation for $m^2(q^2)$
in the infrared limit, i.e, as $q^2\to 0$, where this separation becomes 
unique and totally unambiguous: one assigns to the mass equation all 
contributions that do not vanish as $q^2\to 0$.

It turns out that the equation for  $m^2(0)$ may be derived following two 
rather distinct procedures, 
which eventually express the answer in terms 
of seemingly different quantities.
Roughly speaking, the first procedure, 
which is operationally closer to the standard SDE treatment,    
identifies the contribution to the gluon mass equation as the 
deviation produced to the so-called ``seagull-identity'' by the  
inclusion of the vertex $V$. Interestingly enough, this can be accomplished without 
knowledge of the explicit closed form of the vertex $V$, relying only on its 
general features, most notably its longitudinal nature and 
the Slavnov-Taylor identities that it satisfies.  
The second procedure 
expresses the answer in terms of quantities such as  
``the bound-state wave-function'', denoted by $B_1$, appearing  in the study 
of the Schwinger mechanism from the point of view of the 
actual formation of the required bound state excitations.
This particular quantity 
satisfies a homogeneous Bethe-Salpeter equation,
which, as has been shown recently, admits indeed non-trivial solutions~\cite{Aguilar:2011xe}.

To be sure, 
these two procedures, despite their apparent differences, 
must be ultimately
equivalent.  In this  talk we will argue that this  is indeed the case,
by showing that the two  mass equations obtained coincides in the deep
infrared  limit, thus  providing  an important 
self-consistency check for this entire approach. 
 
\section{Dynamical gluon mass: general concepts and ingredients}  

\noindent Let us start by setting up the notation and reviewing some 
of the most salient features of the dynamical gluon mass generation 
scenario formulated within the PT-BFM framework.
The full gluon propagator 
$\Delta^{ab}_{\mu\nu}(q)=\delta^{ab}\Delta_{\mu\nu}(q)$ in the Landau gauge is defined as
\be
\Delta_{\mu\nu}(q)=- i P_{\mu\nu}(q)\Delta(q^2); \qquad
 P_{\mu\nu}(q) = g_{\mu\nu}- \frac{q_\mu q_\nu}{q^2},
\label{prop}
\ee 
where the scalar factor $\Delta(q^2)$ and the gluon self-energy $\Pi_{\mu\nu}(q)$ are related through
\be
\Delta^{-1}({q^2})=q^2+i\Pi(q^2);\qquad \Pi_{\mu\nu}(q)=P_{\mu\nu}(q)\Pi(q^2).
\label{defPi}
\ee
For later convenience let us also define the {\it inverse} gluon dressing function, $J(q^2)$, as 
\be
\Delta^{-1}({q^2})=q^2 J(q^2);
\label{defJ}
\ee
then, in the presence of a dynamically generated mass, a massive gluon propagator is naturally described in term the natural form of the expression 
\be
\Delta^{-1}_m(q^2) =q^2 J_m(q^2) - m^2(q^2),
\label{defm}
\ee
where the first term corresponds to the ``kinetic term'', or ``wave function'' contribution, 
whereas the second is the (positive-definite) momentum-dependent mass. Notice that the symbol $J_m$ indicates that effectively one has now a mass inside the corresponding expressions: for example, 
whereas perturbatively $J(q^2) \sim \ln q^2$,
after dynamical gluon mass generation has taken place, one has $J_m(q^2) \sim \ln(q^2+m^2)$.

\begin{figure}[!t]
\begin{center}
\includegraphics[scale=0.75]{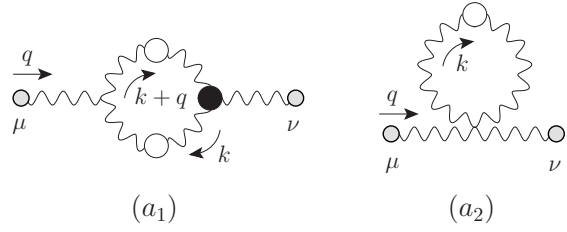}
\caption{\label{gSDE} The ``one-loop dressed'' gluon contribution to the  PT-BFM gluon self-energy. 
White (black) circles denote fully dressed propagators (vertices); 
a gray circle attached to the external legs indicates that they are background gluons. 
Within the PT-BFM framework these 
two diagrams constitute a transverse subset of the full gluon SDE.}
\end{center}
\end{figure} 

The usual starting point of our dynamical analysis is the SDE 
governing the gluon propagator.  
Within the PT-BFM framework that we employ,  
one may safely truncate the SDE series down to its ``one-loop dressed version''  
containing gluonic contributions only, given by the diagrams $(a_1)$ and $(a_2)$ shown in Fig.~\ref{gSDE}~\cite{Aguilar:2006gr,Binosi:2007pi,Aguilar:2008xm}.
In fact, the resulting (approximate) gluon self-energy $\Pi_{\mu\nu}(q)$
is transverse, {\it i.e.}, 
\be
q^{\mu} \Pi_{\mu\nu}(q) = 0. 
\label{transverse}
\ee

The PT-BFM equation for the conventional propagator reads, in this case, 
\be
\Delta^{-1}(q^2){ P}^{\mu\nu}(q) = 
\frac{q^2 {P}^{\mu\nu}(q) + i\,g^2 C_A \sum_{i=1}^{5} A^{\mu\nu}_{i}(q)}{[1+G(q^2)]^2},
\label{sde}
\ee
where, in the Landau gauge,     
\bea
A^{\mu\nu}_{1}(q)&=& \frac{1}{2}\int_k
\Gamma^{(0)\mu}_{\alpha\beta} P^{\alpha\rho}(k)P^{\beta\sigma}(k+q){\bqq}^\nu_{\rho\sigma} \Delta(k) \Delta(k+q),
\nonumber \\
A^{\mu\nu}_{2}(q)&=&  \int_k \! P^{\alpha\mu}(k) \frac{(k+q)^{\beta} \Gamma^{(0)\nu}_{\alpha\beta}}{(k+q)^2}\Delta(k),
\nonumber \\
A^{\mu\nu}_{3}(q)&=& \int_k \! P^{\alpha\mu}(k) \frac{(k+q)^{\beta} {\bqq}^\nu_{\alpha\beta}}{(k+q)^2}\Delta(k),
\nonumber \\
A^{\mu\nu}_{4}(q)&=&-\frac{(d-1)^2}{d}g^{\mu\nu}\int_k \Delta(k), \nonumber \\
A^{\mu\nu}_{5}(q)&=& \int_k \frac{k^{\mu}(k+q)^{\nu}}{k^2 (k+q)^2},
\label{thebees}
\eea
and the $d$-dimensional integral measure (in dimensional regularization) is defined as
$\int_{k}\equiv\frac{\mu^{\epsilon}}{(2\pi)^{d}}\!\int\!\diff^d k.$
The vertex $\bqq$ is the fully-dressed $BQQ$ vertex, connecting a background gluon ($B$) to two quantum gluons ($Q$), which naturally appears in the PT-BFM framework and that has been studied in detail in~\cite{Binosi:2011wi} (see also below); in addition,
\be
\Gamma^{(0)}_{\mu\alpha\beta}(q,r,p) = 
g_{\alpha\beta}(r-p)_\mu +
g_{\beta\mu}(p-q)_\alpha + g_{\alpha\mu}(q-r)_\beta.
\label{treelev}
\ee
Finally, the function $G(q^2)$ represents 
the scalar co-factor of the $g_{\mu\nu}$ component of the special two-point function 
$\Lambda_{\mu\nu}(q)$, defined as 
\bea
\Lambda_{\mu\nu}(q)&=&-ig^2C_A\int_k\!\Delta_\mu^\sigma(k)D(q-k)H_{\nu\sigma}(-q,q-k,k)\nonumber\\
&=&g_{\mu\nu}G(q^2)+\frac{q_\mu q_\nu}{q^2}L(q^2),
\label{Lambda}
\eea
where we have introduced the ghost propagator $D^{ab}(q^2)=\delta^{ab}D(q^2)$, which 
is related to the ghost dressing function $F(q^2)$ through $D(q^2)=  \frac{F(q^2)}{q^2}$
Notice that in the Landau gauge, an important exact (all-order) relation exists, 
linking $G(q^2)$ and  $L(q^2)$ to the 
ghost dressing function $F(q^2)$, namely~\cite{Grassi:2004yq,Aguilar:2009pp}  
\be
F^{-1}(q^2) = 1+G(q^2) + L(q^2);
\label{funrel}
\ee
in addition $G$ coincides with the well-known Kugo-Ojima function~\cite{Grassi:2004yq,Aguilar:2009pp}.

%%%%%%%%%%%%%%%%%%%%%%%%%%%%%%%%%%%%%%%
%   Fig.4
%%%%%%%%%%%%%%%%%%%%%%%%%%%%%%%%%%%%%%%
\begin{figure}[!t]
\begin{center}
\includegraphics[scale=1.3]{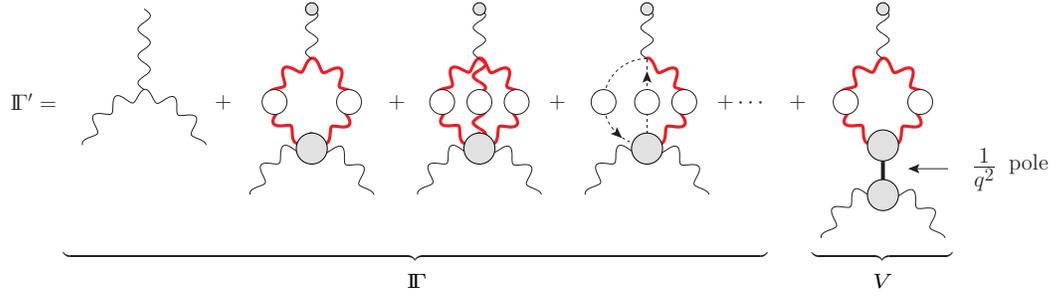}
\end{center}
\vspace{-.5cm}
\caption{\label{Gammaprime}The $\NV$ three-gluon vertex. 
Thick (red online) internal gluon lines indicates gluon propagators with an effective mass.}
\end{figure}
%%%%%%%%%%%%%%%%%%%%%%%%%%%%%%%%%%%%%%%%%%%%%%%%%%%%

The Schwinger mechanism~\cite{Schwinger:1962tn}
is integrated into the SDE of the gluon propagator through the form of the three-gluon vertex~\cite{Jackiw:1973tr,Cornwall:1973ts,Eichten:1974et,Poggio:1974qs}.
In particular,  
a crucial condition for the realization of the gluon mass generation scenario  
is the existence of a special vertex, to be denoted by $V_{\alpha\mu\nu}(q,r,p)$,  
which must be completely {\it longitudinally coupled}, 
{\it i.e.}, must satisfy 
\be
P^{\alpha'\alpha}(q) P^{\mu'\mu}(r) P^{\nu'\nu}(p) V_{\alpha\mu\nu}(q,r,p)  = 0.
\label{totlon}
\ee 
This vertex 
is instrumental for maintaining gauge invariance, 
given that 
the massless poles that it must contain in order to trigger the Schwinger mechanism, 
act, at the same time, as composite, longitudinally coupled Nambu-Goldstone bosons. 
Specifically, in order to preserve the gauge invariance of the theory in the presence of masses, 
the vertex $V_{\alpha\mu\nu}(q,r,p)$ must be added to the 
$BQQ$ (fully-dressed) three-gluon vertex $\bqq_{\alpha\mu\nu}(q,r,p)$, giving rise 
to the new full vertex,  $\NV_{\alpha\mu\nu}(q,r,p)$, defined as~\cite{Aguilar:2011ux}  
\be
\NV_{\alpha\mu\nu}(q,r,p) = \bqq_{\alpha\mu\nu}(q,r,p) +V_{\alpha\mu\nu}(q,r,p).
\label{NV}
\ee
To see in detail how gauge invariance is preserved, notice that when the Schwinger mechanism is turned off, the $BQQ$ vertex alone satisfies the WI
\be
q^\alpha\bqq_{\alpha\mu\nu}(q,r,p)=p^2\bcj(p^2)P_{\mu\nu}(p)-r^2\bcj(r^2)P_{\mu\nu}(r),
\label{STI}
\ee
when contracted with respect to the momentum of the background gluon.
By requiring that 
\be
q^\alpha V_{\alpha\mu\nu}(q,r,p)= m^2(r^2)P_{\mu\nu}(r) - m^2(p^2)P_{\mu\nu}(p) ,
\label{winp}
\ee
we see that,  after turning  the Schwinger mechanism on,  the corresponding WI satisfied by $\NV$ would read  
\bea
q^{\alpha}\NV_{\alpha\mu\nu}(q,r,p) &=& 
q^{\alpha}\left[\bqq(q,r,p) + \NP(q,r,p)\right]_{\alpha\mu\nu}
\nonumber\\
&=& [p^2\bcj (p^2) -m^2(p^2)]P_{\mu\nu}(p) - [r^2\bcj (r^2) -m^2(r^2)]P_{\mu\nu}(r)
\nonumber\\
&=& \Delta^{-1}_m({p^2})P_{\mu\nu}(p) - \Delta^{-1}_m({r^2})P_{\mu\nu}(r) \,,
\label{winpfull}
\eea
which is indeed the identity~(\ref{STI}) satisfied by $\bqq$ with the replacement of the conventional gluon propagators appearing on the rhs by massive propagators: \mbox{$\Delta^{-1} \to \Delta_m^{-1}$}. 
The remaining (more difficult) STIs, triggered when contracting $\NV_{\alpha\mu\nu}(q,r,p)$ 
with respect to the other two legs  
are realized in exactly the same fashion~\cite{Aguilar:2011ux}.

\begin{figure}[!t]
\includegraphics[scale=.72]{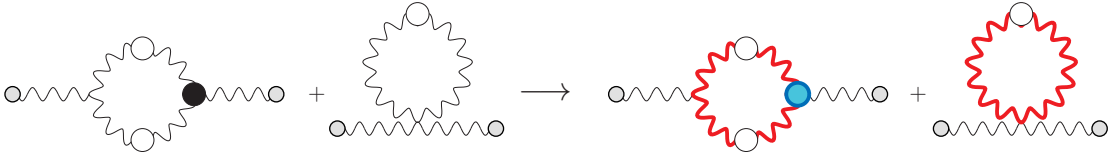}
\caption{\label{mod-gsde}Diagrammatic representation of the gluon one-loop dressed diagrams before and after dynamical gluon mass generation has taken place: the propagators and vertices on the rhs have now become  $\Delta_m$ and $\bqq'$.}
\end{figure}

The next step is to insert $\NV_{\alpha\mu\nu}(q,r,p)$ into the SDE equation satisfied by the 
gluon propagator, thus obtaining the new SDE of Fig.~\ref{mod-gsde}.  
From the resulting SDE one can obtain two separate equations, the first one governing the behavior of $J_m(q^2)$, 
and the second one describing the dynamical mass $m^{2}(q^2)$.
 The general idea is the following: 
the terms appearing on the rhs of the SDE may be separated systematically into two contributions, 
one that vanishes as $q\to 0$ and one that does not; the latter 
contribution must be set equal to the corresponding non-vanishing term on the lhs, namely $-m^{2}(q)$, 
while the former will be set equal to the vanishing term of the lhs, namely $q^{2} J_m(q^2)$.

The exact way how to implement this particular separation is rather elaborate, 
and requires the use of the so-called  
``seagull identity''~\cite{Aguilar:2009ke}
\be
\int_k\! k^2 \Delta'(k)+\frac{d}2\int_k\!\Delta(k)=0,
\label{seagull}
\ee 
where the ``prime'' denotes differentiation with respect to $k^2$. This identity plays a 
a crucial role, forcing all possible quadratic divergences, associated with  
integrals of the type $\int_k\!\Delta(k)$, and variations thereof, to cancel exactly. 
In the case of the 
dimensional regularization that is used throughout, the presence of such integrals would give rise to divergences of the type $m_0^{2} (1/\epsilon)$, where  $m_0$ is the value of the dynamically generated gluon mass at $q^2=0$, 
{\it i.e.}, $m_0 = m(0)$; 
if a hard cutoff $\Lambda$ were to be employed, 
these latter terms would diverge quadratically, as $\Lambda^2$.
The disposal of such divergences would require the introduction in the original 
Lagrangian of a 
counter-term of the form $m^2_0\, A^2_{\mu}$, which is, however, forbidden by the local gauge invariance, which  
must remain intact. Notice that the two types of 
integrals appearing on the lhs  of Eq.~(\ref{seagull}) are individually non-vanishing (in fact, they both diverge); 
it is only when they come in the particular 
combination shown above that they sum up to zero. 

The seagull cancellation implemented by Eq.~(\ref{seagull}) is a point of paramount importance, 
because it further re-enforces the   
fundamental assertion that 
the gluon mass generation, when implemented correctly,  
{\it is absolutely compatible with the underlying BRST symmetry}.
Indeed, in the mass generation 
picture advocated in a series of recent articles~\cite{Aguilar:2008xm,Aguilar:2011ux,Aguilar:2009nf,Aguilar:2010zx} 
{\it the Lagrangian of the Yang-Mills theory (or that of QCD) is never altered};  
therefore, the only other possible ways of violating the gauge (or BRST) symmetry 
would be ({\it i}) 
by not respecting, at some intermediate step, some of the WIs and STIs satisfied by the 
Green's functions involved; for example, in the conventional SDE formulation, a naive truncation would 
compromise the transversality of the resulting gluon self-energy, {\it i.e.}, the fundamental transversality condition~(\ref{transverse}) would be no longer valid; and/or ({\it ii})  by introducing seagull-type divergences, of the 
type mentioned above. Evidently, neither ({\it i})  nor ({\it ii})  happens, thanks to the powerful 
field-theoretic properties encoded into the PT-BFM formalism that we employ.

Returning to the gluon SDE, 
a rather elaborate analysis~\cite{Aguilar:2011ux}  gives rise to two coupled integral equations, 
one for $J_m(q)$ and one for 
the momentum-dependent gluon mass, 
of the generic type
\bea
J_m(q^2) &=& 1+ \int_{k} {\cal K}_1 (q^{2},m^2,\Delta_m),
\nonumber\\
m^{2}(q^2) &=&  \int_{k} {\cal K}_2 (q^{2},m^2,\Delta_m).
\label{separ}
\eea
such that $q^{2} {\cal K}_1 (q^{2},m^2,\Delta_m) \to 0$, as $q^{2}\to 0$, 
whereas ${\cal K}_2 (q^{2},m^2,\Delta_m)\neq 0$ in the same limit,
precisely because it includes the term $1/q^2$ contained inside $V_{\alpha\mu\nu}(q,r,p)$. 

There is a relatively straightforward way to derive the closed form of the gluon mass equation in the limit $q^{2}\to 0$, 
without knowledge of the explicit form of $V$;  all one needs is to postulate its existence and the properties mentioned above,
most notably the identities~(\ref{totlon}) and~(\ref{winp}). 
In order to do that, we start by noticing that, 
since the vertex $\bqq'$ satisfies the identity~(\ref{winpfull}), 
the gluon self-energy is transverse {\it even in the presence of masses}; as a result, restoring the Lorentz structure 
into the second equation of (\ref{separ}), we have that 
\be
m^{2}(q^2) P_{\mu\nu}(q)  = P_{\mu\nu}(q) \int_{k} {\cal K}_2 (q^{2},m^2,\Delta_m).
\ee
Now, due to its longitudinal nature, 
the vertex $V$  can only furnish the part on the rhs 
proportional to $q_{\mu} q_{\nu}/q^2$. The question is, where will the $g_{\mu\nu}$ part come from.
A bit of thought reveals that this term can only emerge as a deviation from the  seagull identity, 
which enforces the masslessness of the gluon when $V=0$. 
In fact, this identity, given its nature and function,
can only operate among terms that 
are proportional to $g_{\mu \nu}$; this is so because the basic seagull contribution 
proportional to $\int_k\!\Delta(k)$ stems entirely from graph $(a_2)$, 
which has no momentum dependence, {\it i.e.}, it can only be proportional to  $g_{\mu \nu}$.  

Specifically, the required contribution proportional to 
$g_{\mu \nu}$ stems from the term 
\be
\int_k\!k_\mu k_\nu\Delta(k)\Delta(k+q)\frac{(k+q)^2J_m(k+q)-k^2J_m(k)}{(k+q)^2-k^2}=
g_{\mu\nu}{C}_1(q^2)+\frac{q_\mu q_\nu}{q^2}{C}_2(q^2),
\label{theCs}
\ee
which originates from $A^{\mu\nu}_{1}(q)$.
Then, since for any function $f(k^2)$ one has the result
\be
\int_k\!\cos^2\theta f(k^2)=\frac1d\int_k\!f(k^2),
\label{costheta-rel}
\ee
it is easy to demonstrate that 
\bea
{C}_1(0) &=& \int_k k^2 \Delta^2(k) [k^2 J(k)]'
\nonumber\\
&=& \int_k k^2 \Delta^2(k) [\Delta^{-1}(k)+m^2(k)]'
\nonumber\\
&=& \int_k k^2 \Delta^2(k) [m^2(k)]' -\int_k k^2  \Delta'(k).
\label{C10}
\eea
The first term on the rhs forms part of $ g_{\mu \nu} {\cal K}_2 (q^{2},m^2,\Delta_m)$, 
while the second, after taking all relevant multiplicative 
factors correctly into account, cancels against the contribution of graph $(a_2)$, 
by virtue of the seagull identity~(\ref{seagull}). 
This basic observations can be generalized to include 
the additional structures coming from the remaining terms in~(\ref{thebees}), 
and in particular  $A^{\mu\nu}_{3}(q)$, whose net effect is the one of reducing by one power the $1+G$ 
factor appearing in the denominator of Eq.~(\ref{sde})~\cite{Aguilar:2011ux}.

Then, restoring all relevant factors, and using the fact that in 4 dimension $L(0)=0$~\cite{Aguilar:2009pp} so that by virtue of Eq.~(\ref{funrel}) one can trade the $1+G$ combination for the inverse of the ghost dressing function, one finally obtains, in Euclidean space,
\bea
m^2(0)&=&\frac{3}{2} g^2 C_A F(0)\int_k\!k^2[m^2(k)]'\Delta^2(k)\nonumber \\
&=&-3g^2C_A F(0) \int_k\!m^2(k)\Delta(k)\left[k^2\Delta(k)\right]'.
\label{me-qtozero}
\eea 
and, after carrying out the angular integration, the final equation
\be
m^2(0) = -\frac{3C_A}{8\pi}\alpha_s F(0) \int_0^\infty\!\diff y\,m^2(y) [y^2\Delta^2(y)]',
\label{m20}
\ee 
where \mbox{$\alpha_s=g^2/4\pi$} and \mbox{$y=k^2$} 
(and therefore the ``prime'' indicates now derivatives with respect to~$y$).  

%%%%%%%%%%%%%%%%%%%%%%%%%%%%%%%%%%%%%%%%%%%%%%%%%%%%%%%%%%%%%%%%%%%%%%%%%%
%             Fig. derivative of the gluon dressing squared
%%%%%%%%%%%%%%%%%%%%%%%%%%%%%%%%%%%%%%%%%%%%%%%%%%%%%%%%%%%%%%%%%%%%%%%%%%%%
\begin{figure}[!t]
\hspace{-1cm}\includegraphics[scale=.96]{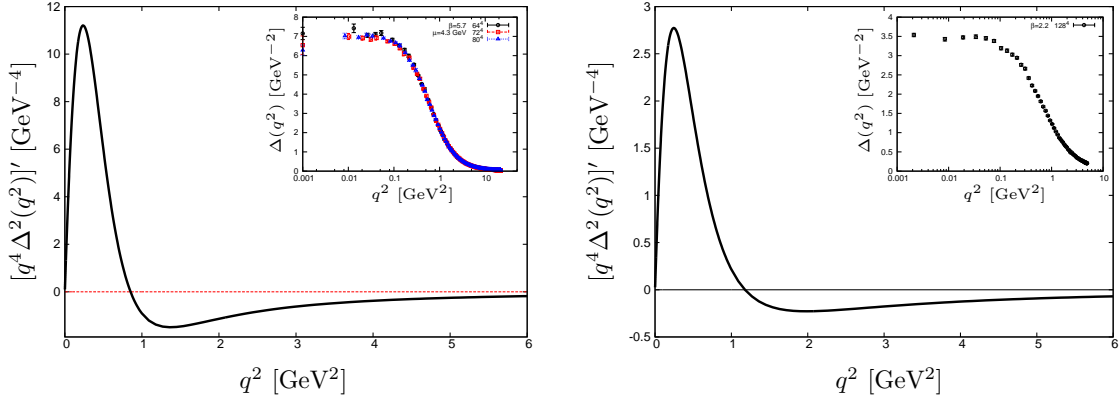} 
\caption{\label{gprop} The kernel $[q^4\Delta^2(q^2)]'$ 
obtained from the $SU(3)$ (left) and $SU(2)$ (right) lattice data. In the inset we show the corresponding lattice results for the gluon propagator, renormalized  at $\mu=4.3$ GeV and $\mu=2.2$ GeV respectively.}
\end{figure}
%%%%%%%%%%%%%%%%%%%%%%%%%%%%%%%%%%%%%%%%%%%%%%%%%%%%%%%%%%%%%%%%%%%%

It turns out that the specific form of the mass equation~(\ref{m20})  introduces a 
non-trivial constraint on the precise behavior that $\Delta$ must display in the region between \mbox{(1-5) $\rm GeV^2$}. 
Specifically, in order for the gluon mass to be positive definite,
 the first derivative 
of the quantity $ q^2 \Delta(q^2)$ (the ``gluon dressing function'')
must furnish a sufficiently {\it negative} 
contribution in the aforementioned range of momenta. Interestingly enough, the 
$\Delta$ obtained from the lattice, shown in Fig.~\ref{gprop}, has indeed this particular property; 
in the plots shown in Fig.~\ref{gprop}  we clearly 
see that both derivatives change their sign in the 
intermediate momenta region, which 
constitutes precisely the required behavior. 
This is to be contrasted 
to what happens, for example, in the case of a simple massive propagator $1/(q^2+m^2)$  
or with the Gribov-Zwanziger propagator $q^2/(q^4+m^4)$ (with $m$ constant); 
the derivatives of the corresponding dressing functions, 
$q^2/(q^2+m^2)$ and $q^4/(q^4+m^4)$, respectively,  
are positive in the entire range of (Euclidean) momenta,
thus excluding, in this context, the possibility of a positive-definite gluon mass.

\section{Schwinger mechanism in the bound-state language: a self-consistency check}

%%%%%%%%%%%%%%%%%%%%%%%%%%%%%%%%%%%%%%%%%%%%%%%%%%%%%%%
\begin{figure}[!t]
\center{\includegraphics[scale=1]{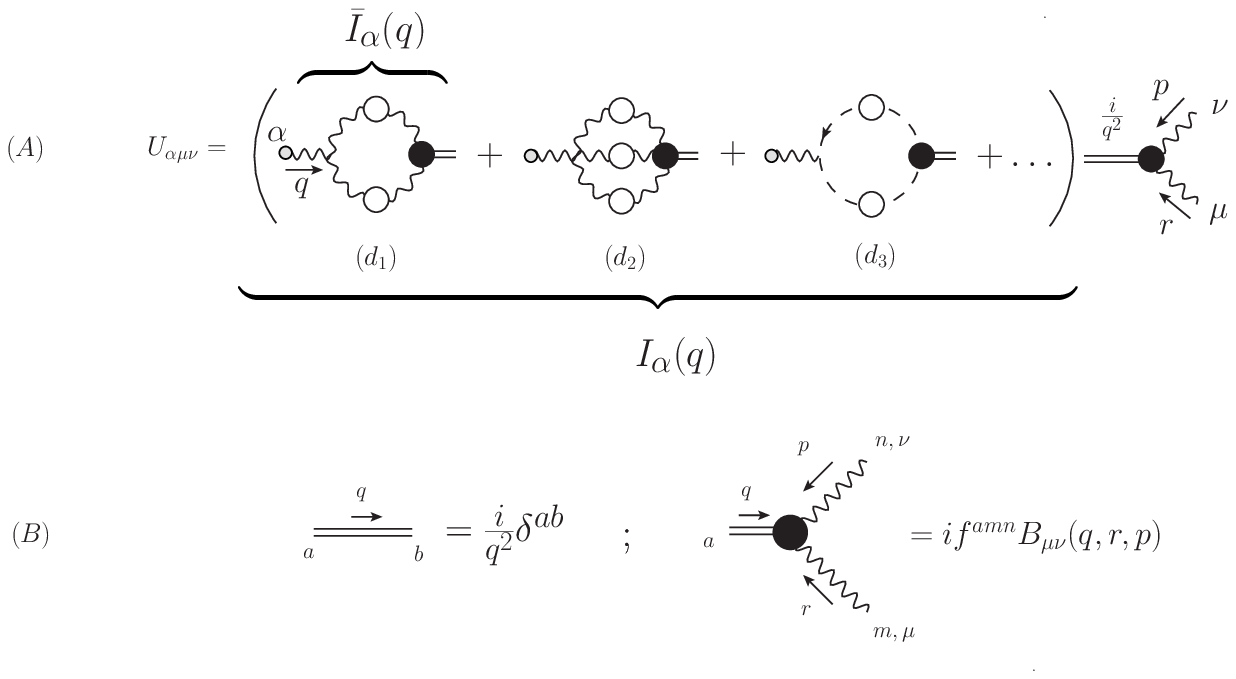}}
\caption{\label{Uexpansion} ({\it A}) The vertex $U_{\alpha\mu\nu}$ is composed of three main ingredients: 
the transition amplitude, $I_{\alpha}$, which 
mixes the gluon with a massless excitation, 
the propagator of the  massless excitation, and the massless excitation--gluon--gluon vertex. 
({\it B}) The Feynman rules (with color factors included) for ({\it i}) the propagator of the massless excitation 
and ({\it ii}) the ``proper vertex function'', or, ``bound-state wave function'', $B_{\mu\nu}$.}  
\end{figure}
%%%%%%%%%%%%%%%%%%%%%%%%%%%%%%%%%%%%%%%%%%%%%%%%%%%%%%%

The key assumption when invoking the Schwinger mechanism in Yang-Mills theories, 
such as QCD, is that the required  poles   may be produced 
due to purely dynamical reasons; specifically, one assumes that, for sufficiently 
strong binding, 
the mass of the appropriate bound state may be reduced to zero~\cite{Jackiw:1973tr,Cornwall:1973ts,Eichten:1974et,Poggio:1974qs}.
It is precisely these poles that constitute the main ingredient of the special vertex $V$.
Specifically, all terms appearing in this vertex are proportional to $1/q^2$, $1/r^2$,  $1/p^2$, 
and products thereof, and can be separated in two distinct parts, according to~\cite{Aguilar:2011xe}
\be
V_{\alpha\mu\nu}(q,r,p) =  U_{\alpha\mu\nu}(q,r,p) + R_{\alpha\mu\nu}(q,r,p).
\label{VRU}
\ee
The $R$ term contains structures proportional to $r_{\mu}/r^2$ and/or $p_{\nu}/p^2$, which vanish when contracted  with the product $P_{\mu'\mu}(r)P_{\nu'\nu}(p)$ coming from the Landau gauge gluon propagators, and therefore are not relevant for us. On the other hand, the $U$ term can be written as~[see Fig.~\ref{Uexpansion} ({\it A})]
\be
{U}_{\alpha\mu\nu}(q,r,p) = iI_{\alpha}(q)\left(\frac{i}{q^2} \right) \ff_{\mu\nu}(q,r,p), 
\label{VwB}
\ee
where the factor ${i}/{q^2}$ represents the propagator of the scalar massless excitation, while the nonperturbative quantity 
\be
\ff_{\mu\nu}(q,r,p) = B_1g_{\mu\nu}+B_2q_{\mu}q_{\nu}+B_3p_{\mu}p_{\nu}+B_4r_{\mu}q_{\nu}+B_5r_{\mu}p_{\nu},
\ee
is the effective vertex 
describing the interaction between 
the massless excitation and two gluons [Fig.~\ref{Uexpansion} ({\it B})]. Finally,   
$I_{\alpha}(q) = q_{\alpha} I(q)$ is the (nonperturbative) transition 
amplitude introduced in Fig.~\ref{Uexpansion}, allowing the 
mixing between a gluon and the massless excitation\footnote{Notice that  
we do  not absorb the extra factor of ``i'' coming from the Feynman rule in Fig.\ref{Uexpansion} 
into the definition of the transition function $I_{\alpha}$, as was done in~\cite{Aguilar:2011xe};
this definition has the advantage that the function $I(q)$ is real in the Euclidean space.}. Ii is important to notice that, 
due to Bose symmetry with respect to the quantum legs, {\it i.e.}, $\mu\leftrightarrow\nu$ and $p\leftrightarrow r$ 
one has $B_{1,2}(q,r,p)=-B_{1,2}(q,p,r)$, and therefore these two form factors vanish in the limit. $q\to0$.

The WI~(\ref{winp}) furnishes an  
exact relation between the dynamical gluon mass, the transition amplitude at zero momentum transfer, 
and the form factor $B_1$. 
Specifically, contracting both sides of the WI with 
two transverse projectors, one obtains,
\be
P^{\mu'\mu}(r) P^{\nu'\nu}(p) q^\alpha V_{\alpha\mu\nu}(q,r,p) = [m^2(r)-m^2(p)] P^{\mu'}_\sigma(r) P^{\sigma\nu'}(p).
\ee
On the other hand, contracting the full expansion of the vertex (\ref{VwB}) by the same transverse projectors
and next contracting the resulting expression with the momentum of the background leg, we get 
\be
q^\alpha P^{\mu'\mu}(r) P^{\nu'\nu}(p) V_{\alpha\mu\nu}(q,r,p) = -I(q)[B_1g_{\mu\nu}+B_2q_\mu q_\nu] P^{\mu'\mu}(r) P^{\nu'\nu}(p),
\ee
Thus, equating both results, one arrives at the exact relations
\be
I(q) B_1(q,r,p) = m^2(p)-m^2(r);\qquad B_2(q,r,p) = 0 . 
\label{factor1}
\ee
Finally, carrying out the Taylor expansion around $q=0$ of both sides of 
the first relation in~(\ref{factor1}), and using the fact that the form factors $B_{1,2}$ vanish in this limit, we arrive at (Minkowski space)
\be
[m^2(p)]' =  - I(0) B'_1(p), 
\label{massrelation}
\ee
where we have defined
\begin{equation}
B'_1(p)=B_1'(0,-p,p) \equiv \, \lim_{q\to 0} \left\{ \frac{\partial B_1(q,-p-q,p)}{\partial\, (p+q)^2} \right\},
\label{Der}
\end{equation}
and have implicitly assumed that $I(0)$ is finite.

%%%%%%%%%%%
%   Fig.5
%%%%%%%%%%%%%%%%%%%%%%%%%%%%%%%%%%%%%%%%%%%%%%%%%%%%%%%%%%%%%%%%%
\begin{figure}[t]
\center{\includegraphics[scale=0.55]{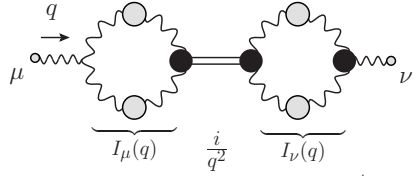}}
\caption{The ``squared'' diagram.}
\label{Square}
\end{figure}
%%%%%%%%%%%%%%%%%%%%%%%%%%%%%%%%%%%%%%%%%%%%%%%%%%%%%%%%%%%%%%%%%%%%%%%%%%%%%%%%%%%%%%%%

We will next approximate the transition amplitude $I_{\alpha}(q)$, connecting the gluon with the 
massless excitation,    
by considering only diagram $(d_1)$ in Fig.~\ref{Uexpansion}, corresponding to the 
gluonic ``one-loop dressed'' approximation;  
we will denote the resulting expression by ${\bar I}_{\alpha}(q)$. 
In the Landau gauge, the amplitude ${\bar I}_{\alpha}(q)$ reads 
\be
{\bar I}_{\alpha}(q) =  \frac{1}{2} i\, C_A 
\int_k \Delta(k) \Delta(k+q) \Gamma_{\alpha\beta\lambda}^{(0)} P^{\lambda\mu}(k) P^{\beta\nu}(k+q) \ff_{\mu\nu}(-q,-k,k+q),
\label{Iappr1}
\ee
where the origin of the $1/2$ factor  is combinatoric, 
and $\Gamma_{\alpha\beta\lambda}^{(0)}$ is the tree-level vertex of Eq.~(\ref{treelev}). 

%A a bit of algebra reveals finally that, in Minkowski space,~\cite{Aguilar:2011xe}
%\begin{equation}
%{\bar I}(0) = \frac{3}{2} i\, C_A\int_k k^2\Delta^2(k)\ff'_1(k).
%\label{zero}
%\end{equation}

At this point we are in the position to determine how the transition function is related to the dynamically generated gluon mass which originates from the inclusion of the vertex $V$ in the corresponding gluon SDE. Indeed, at the level of approximation employed here we find that the contribution to the gluon self energy is   
\bea
A^{\mu\nu}_1(q)|_{\chic V}  &=& 
\frac{1}{2} g^2 C_A \int_k \Delta(q+k)\Delta(k) 
\Gamma^{(0)\mu}_{\alpha\beta}P^{\alpha\rho}(k)P^{\beta\sigma}(k+q) U^\nu_{\rho\sigma}
\nonumber\\
&=& 
-g^2 \left[\frac{1}{2} C_A \int_k \Delta(q+k)\Delta(k) 
\Gamma^{(0)\mu}_{\alpha\beta}P^{\alpha\rho}(k)P^{\beta\sigma}(k+q)
\ff_{\rho\sigma} \right] \left(\frac{i}{q^2} \right){\bar I}_{\nu}(-q) 
\nonumber\\
&=& -i \frac{q^{\mu}q^{\nu}}{q^2} \,g^2  {\bar I}^{\,2} (q),
\label{A1v}
\eea
where we have used Eq.~(\ref{VwB}) (with $I_{\nu} \to {\bar I}_{\nu}$), 
together with the property \mbox{${\bar I}_{\nu}(-q) = -{\bar I}_{\nu}(q)$} as well as
Eq.~(\ref{Iappr1}). Notice that as anticipated this contribution is purely longitudinal. 

Since the inclusion of V in $A_3$ 
has the same effect previously found, namely to effectively remove one power of $(1+G)$ in the denominator\footnote{It is precisely the inclusion of this term that account for the difference of Eq.~(\ref{me-geneuc}) with respect to the analogous equation (3.38) of~\cite{Aguilar:2011xe}.}, we obtain
the positive-definite result
\be
m^2(0)= g^2 F(0) {\bar I}^{\,2} (0).
\label{me-geneuc}
\ee

Of course, for the PT-BFM framework to be self-consistent, the two infrared equations for the gluon mass ought to coincide. To see that this is indeed the case, consider again the original mass equation~(\ref{me-qtozero}) , and let's insert the relation~(\ref{massrelation}) -- at the same level of approximation used in the derivation of~(\ref{me-geneuc}), {\it i.e.}, with the replacement $I(0)\to\bar I(0)$ -- into the integral in the first line, to get
\bea
m^2(0)&=&\frac{3}{2} g^2 C_A F(0)\int_k\!k^2[m^2(k)]'\Delta^2(k)\nonumber \\
&=& g^2 F(0) 
I(0) \left[-\frac{3}{2} g^2 C_A \int_k\!k^2 \Delta^2(k) \ff'_1(k)\right].
\label{last}
\eea
On the other hand, a a bit of algebra reveals  that, in Euclidean space, on has ~\cite{Aguilar:2011xe}
\begin{equation}
{\bar I}(0) = \frac{3}{2} i\, C_A\int_k k^2\Delta^2(k)\ff'_1(k),
\label{zero}
\end{equation}
so that we indeed~(\ref{last}) gives rise to the relation
\be
m^2(0)=g^2 F(0) {\bar I}^{\,2} (0),
\ee 
which is none other than Eq.~(\ref{me-geneuc}). 

\section{Conclusions}

\noindent In this work we have  presented a 
highly non-trivial self-consistency check 
related to the integral 
equation describing the dynamics of the gluon mass 
in the deep infrared. 

We started by showing how relation~(\ref{me-qtozero}) can  be
derived  only by  assuming the  existence of  the special  vertex $V$,
without  providing  any information  either  on  its own  diagrammatic
structure,  nor  on  its   composition  in  terms  of  other  dynamical
quantities,  such  as the  bound-state  wave  function  $B$. Next,  we
presented  an  alternative  formulation, where all  the  intricate
dynamical ingredients that  trigger the Schwinger mechanism, subject
to the  constraints imposed by  gauge invariance, were  furnished. Both
methods give rise  to the same equation, proving  that they represent
nothing but different facets of the same underlying dynamics.

In the  future it  would be particularly  important to  generalize the
consistency check  described to the  full momentum range,  {\it i.e.},
extend it  to the full  dynamical equation governing the  evolution of
the  mass  $m^2(q^2)$  and  not   only  to  the  deep  infrared  limit
$m^2(0)$. In addition, one should go beyond the one-loop approximation
by including the  missing two-loop dressed gluon diagrams  in both the
SDE as well  as the bound-state analysis. Work  along these directions
is already in progress.

\newpage 

\noindent {\it Acknowledgments} 

\noindent The research of J.P. was supported by the European FEDER and  Spanish MICINN under 
grant FPA2008-02878.

\end{document}